\documentclass{raa_new}

\usepackage{graphicx,times}
\usepackage[numbers,super]{natbib}
\usepackage{CJK,color}
\usepackage{url}
\usepackage{fancyhdr}

\def\cite{\citep}
\def\psrsoft{{\sc PSRSoft}}

\makeatletter
\g@addto@macro\@verbatim\small
\makeatother

\begin{document}
\makeatletter
\def\@cite#1#2{\textsuperscript{[{#1\if@tempswa  #2\fi}]}}
\makeatother
\begin{CJK*}{GBK}{song}


   \volnopage{Vol.9 (2012) No.3, 219--229}
   \setcounter{page}{219}          
   \author{M. J. Keith
   }
   \institute{CSIRO Astronomy \& Space Science, ATNF, P.O. Box 76, Epping 1710, NSW Australia
   }
   \date{Received~~2012 month day; accepted~~2012~~month day}    
  \ \ \abstract{Searching for radio pulsars typically requires a bespoke software pipeline to
efficiently make new discoveries. In this paper we describe the search process, provide a
tool for installing pulsar software, and give an example of a pulsar search.
   }

\title{Installation and Use of Pulsar Search Software}
   \rhead{M. J. Keith:~ Installation and Use of Pulsar Search Software}
   \lhead{}

   \maketitle
\end{CJK*}
%
%
\section{Introduction}           
Searching for radio pulsars is a complex and time-consuming task, however we are continually motivated to make further discoveries by the important and often unique science that flows from the study of these objects.
The first pulsar discoveries were made by detecting individual pulses on pen-chart recorders, however even with large telescopes only the brightest and closest pulsars can be detected through this method.

There are two aspects to the pulsar signal that distinguish it from other astronomical signals and man-made radio frequency interference (RFI).
Firstly, the constant spin period of the pulsar leads to a very regular train of pulses with periods ranging from 0.001 to 10$\,$s.
Secondly, free electrons in the interstellar medium cause the pulse to be dispersed, causing a group delay at frequency $\nu$,
\begin{equation}
\tau(\nu) = \nu^{-2}  \frac{e^2}{2{\rm\pi} m_ec} \int_{path} \!\!\!\!\!\! n_e (l) \mbox{d}l ,
\end{equation}
where $n_e$ is the electron density along the line of sight to the pulsar.
We term the path integral the dispersion measure, DM, and give it in units of cm$^{-3}$pc.
Although this adds complexity to the analysis of pulsar data, it also provides a clear distinction between astronomical and terrestrial signals.
Therefore, if we are to discover new pulsars, we must develop a strategy for detecting dispersed and highly periodic radio pulses.

In this paper we will discuss the general pulsar search strategy (Section 1), describe how to install the required pulsar search software using \psrsoft\ (Section 2) and give an example of how to process an observation from the Parkes Radio Telescope (Section 3).

\section{The Pulsar Search Strategy}
\begin{figure}[htb]
\begin{center}
\includegraphics[width=4cm]{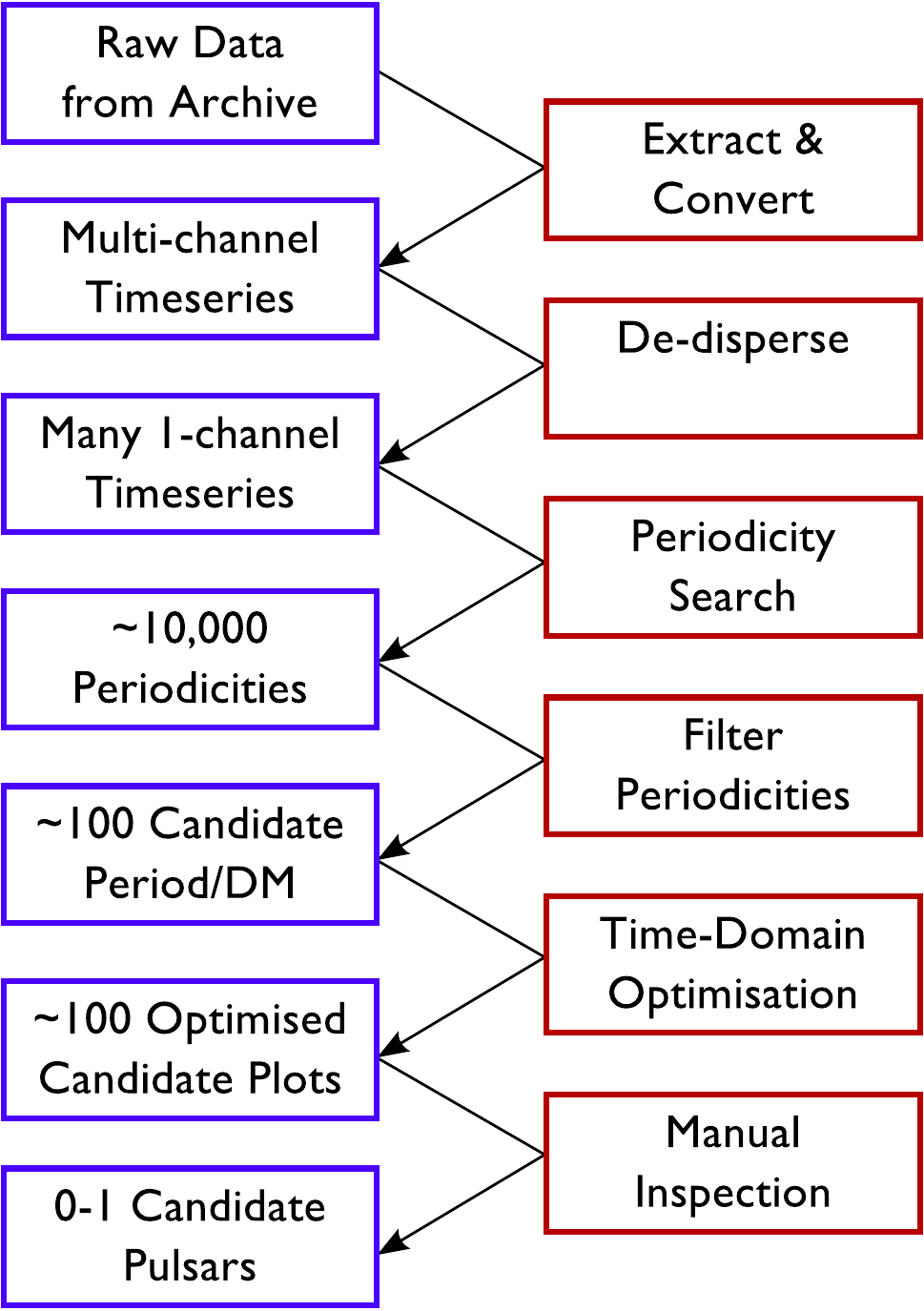}
\caption{
\label{searchoverview}
Schematic overview of the pulsar search process. Data products are shown on the left and software processes are shown on the right.
}
\end{center}
\end{figure}

The search for radio pulsars is the search for periodic, dispersed, radio signals on the sky.
Therefore, this defines the parameter space over which we must search: position on the sky, DM, and pulse period.
Additionally, in order to increase sensitivity to binary pulsars for which the pulse period varies due to the Doppler effect, we may need to include additional search parameters such as acceleration and higher order terms.
Luckily, these parameters are generally independent, and so we can form them into a hierarchical structure where we loop over each parameter in turn.
Figure \ref{searchoverview} shows a schematic overview of the pulsar search process, with the following steps:
\begin{itemize}
\item {\bf Extraction} --- this covers obtaining the relevant data file from the telescope data archive, and converting the data to a compatible data format.
\item {\bf De-dispersion} --- here we change our single file with multiple frequency channels to a large number of time-series each correcting for a different value of DM.
\item {\bf Periodicity search} --- each of the de-dispersed time-series must be searched for periodic signals, or for bright individual pulses.
\item {\bf Filter candidates} --- periodicities detected in each of the time-series are grouped together to produce the final candidate listing.
\item {\bf Optimisation} --- finally, the original data can be `folded' to optimise the period and DM, and generate the diagnostic plots used to distinguish good candidate pulsars from RFI.
\end{itemize}

\subsection{Data Extraction}
A typical pulsar search data file will be a recording of a single position on the sky, with 64 to 1024 frequency channels sampled every 32 to 256 $\mu$s.
Occasionally, multiple positions will be recorded into the same file, e.g. when using a multi-beam receiver such as that used in the Parkes Multibeam Pulsar Survey \cite{mlc+01}, however even in this case software is provided to convert the data a single observation per file.
Unfortunately different instruments often record data in different formats, and converting the data to a usable format may require some inside knowledge of the instrument.
Some effort is being made, however, to adopt the {\sc psrfits} data format across many different instruments, and large databases such as the Parkes Pulsar Data Archive have converted all the data to {\sc psrfits} \cite{hmm+11}.

\subsection{De-Dispersion}
As the pulsar signals are dispersed by the interstellar medium, the pulse arrival time is delayed by different amounts at each frequency and the delay is proportional to $\nu^{-2}$.
To counteract this, the telescope signals are recorded in many narrow frequency channels, and the effects of dispersion are reversed digitally.
This process, known as de-dispersion, is simply a matter of applying a correcting time offset to each frequency channel and summing the channels.
However for search observations the amount of dispersion suffered by the potential source is unknown, and this process must be repeated with many trial values of DM.
This produces a large number of single-channel timeseries, for which the subsequent processing can be performed independently.
For large searches, it may be beneficial to distribute these time-series to many different compute cores, allowing for the search to be carried out in parallel.

\subsection{Periodicity Search}
The Fourier transform is an ideal way to detect the constant periodic signals produced by the majority of known pulsars. The Fourier transform is a decomposition of a data series into a sum of complex exponentials (i.e. sine waves). In this way an infinite sinusoidal wave will form a single infinitely narrow spike at the frequency corresponding to the period of the signal. More generally any periodic signal will form a peak at the fundamental frequency, plus a series of harmonics, with magnitudes depending on the shape of the input signal. By transforming the input time-series in this way, any pulsar signals in the data should be among the strongest signals in the Fourier spectrum, even if they are undetectable in the original time-series.

Whilst the fundamental frequency of the pulsar is the strongest signal produced in the Fourier spectrum, pulsars with narrow pulses will have a lot of power `lost' in the harmonics.
By using a technique known as harmonic summing, it is possible to regain some of this `lost' power.
This is achieved by stretching the power spectrum by a factor of two, and summing it with the un-stretched power spectrum. In this way, the power from the second harmonic will be added on top of the fundamental power, increasing the signal by more than the $\sqrt{2}$ increase in noise (The increase depends on the pulse shape). One can then select the best signals from this new spectrum. This can be repeated to add more harmonics. However since the noise is also increased upon every addition, the S/N will reach a maximum at a harmonic fold related to the pulse width.
Therefore it is important to keep the results from each harmonic fold used, and the harmonic fold that a pulsar was detected strongest in can give an indication of the width of the pulsations.
Since the harmonic summing uses the power spectrum, it is possible that the process will boost the S/N for signals that are not truly harmonically related.
Alternative methods of coherently summing harmonics can be used to identify the signals that are truely harmonically related.

The principal algorithm used to perform Fourier domain searches is the Fast Fourier Transform \cite{ct65}. This algorithm computes a discrete Fourier transform in the order of $N \log_2 N$ operations, for transform length $N$. The improvement over a traditional DFT algorithm is immense, especially when the number of data points is large. For example, with the number of data points at $2^{23}$, a typical observation length, the computation time is reduced by a factor of 105.
Recent work has looked to accelerate this processing even further through the use of the powerful graphics processing units of modern computers.
By using the highly parallel processing architecture of these chips, the time for all parts of the search may be reduced dramatically.

\subsubsection{Binary Motion Compensation}
Binary pulsars have an apparent period drift due to the Doppler effect.
If the binary period is of the same order as the observation time, the changing period will mask the signal in the Fourier domain.
This is due to the power being distributed in different frequency bins.

These effects can be effectively removed from a time-series before processing, however to completely remove all the effects the binary parameters must be known.
There are unfortunately too many parameters to perform a search for all possible orbits in a reasonable time scale.
By expanding the orbital motion as a polynomial, it is possible to approximate the orbit by using acceleration and jerk\footnote{Jerk is used in this document as a convenient contraction for `rate of change of acceleration'.}.
By using this approximation, searching for binary pulsars can be done with only one or two extra parameters.
There are three typical methods for removing this acceleration from the search process, either by re-sampling the time-series \cite{mk84,eklk11}, using incoherent methods \cite{fsk+04}, or by using matched filters in the Fourier domain \cite{rem02}.

The `simplest' method is to re-sample the time series with a new sample rate that varies with time, corresponding to a particular value of acceleration and jerk.
This effectively reverses out the binary motion and so the corrected time-series can then be searched using exactly the same methods as the uncorrected time-series.
Since the optimal acceleration and jerk are not known beforehand, it is necessary to search over a large range of parameters to get the optimal S/N.

Alternatively, one can search for acceleration in the frequency domain.
The number of bins which the pulse power is spread across for a given orbital acceleration is proportional to the observation time.
Therefore this spread can be reduced by splitting the observation into $n$ smaller segments.
The signal strength in each part is however reduced by a factor of $1/n$, however by adding the power-spectra along lines of constant acceleration we can recover $1/\sqrt{n}$ of the original signal strength ratio.
Although this incoherent method is less sensitive, it is very fast.

In order to recover all the original signal strength in the frequency domain, we can use matched filters.
When using matched filters, the time-series is Fourier transformed as normal, however the resulting spectra are then analysed by convolving with the known response from each trial acceleration and jerk.
This methods does not require repeated Fourier transformations, and so can be quicker than the time domain methods.
Additionally, since this method searches over a parameter space in the frequency domain, where the search is actually done, the parameter space is never over-sampled.

\subsubsection{Single Pulse Search}
The flux density of individual pulses can vary considerably, and it is possible to detect many pulsars by the detection of individual strong pulses. In addition to this, there are classes of pulsar-like objects that have been shown to show so few total pulses that standard periodicity searches cannot detect them \cite{mll+06}.
The detection of single pulses relies on the pulses being highly dispersed relative to terrestrial burst emission. Pulsars can be detected by one or more dispersed pulses that are significant above the general noise, and not associated with any un-dispersed terrestrial interference\cite{cm+03}.

\subsection{Candidate Filtering}
The periodicity search typically detects many hundreds of signals for each trial DM.
Fortunately, many of these are harmonics of stronger signals, and many signals are detected at multiple DM trials.
To reduce this to a manageable number of candidates, the signals are grouped together by period, keeping information about which DM trials it was detected in.
This method can typically reduce the number of candidates to $\sim 100$ per observation.

\subsection{Time Domain Optimisation}
Once we have a small number of candidates, we can return to the original timeseries and `fold' the data at the candidate period.
This process steps through the time-series adding each sample to an appropriate output bin depending on the computed pulse phase at that moment.
To fold data correctly requires more accuracy in the period than is provided by the finite width bins of a Fourier analysis.
Therefore it is common to split the time-series into a number of sub-integrations, and fold those separately.
With the shorter folds, the pulse will drift less in phase and therefore have less broadening.
This also allows for the measurement of the variation of signal over the observation, providing useful clues for identifying pulsars.

Summing the sub-integrations is equivalent to folding the entire time-series.
Shifting each sub-integration in phase before summing can compensate for errors in the initial folding period.
By shifting and adding the sub-integrations in this way, one can perform an approximate folding of the time-series at a number of different periods in a computationally efficient manner.
In a similar fashion, using quadratic or cubic shifts with time can compensate for orbital acceleration and jerk.
When the offsets correspond to the correct pulse period, etc. the power from the pulse should all fall in a minimum set of bins.
The correction of the sub-integrations increases the effective signal strength, and therefore the S/N can be maximised.
The way that the signal strength varies with period shows how well aligned in phase the pulses are, and so this information is useful when distinguishing highly accurate pulsar signals from man-made interference.

\subsection{Candidate Selection}
Finally, we have produced candidate pulsars ready for inspection.
Unfortunately, pulsar search pipelines often produce hundreds of candidates per observation, which often precludes manual inspection of each candidate.
A simple method of selecting candidates for evaluation is to sort all candidates by S/N, however this often returns many instances of man-made RFI.
More recently, graphical selection tools \cite{kel+09} or artificial neural nets \cite{emk+10} have been employed to allow for a more efficient selection of pulsar candidates.

\section{Installing Software with \psrsoft}
Although the pulsar search process is conceptually simple, it can take a lot of time and effort to create a efficient and effective pipeline.
Fortunately, many tools have already been developed for pulsar searching, and new algorithms are frequently being developed.
A typical search pipeline may, however, depend on many different software packages.
To help with installing this large variety of software, a software package installation tool \psrsoft\ is provided.

This section gives an overview of how to set up a pulsar search system with \psrsoft.
Before you install \psrsoft\ you must ensure that you have a {\sc python} interpreter and a C, C++ and FORTRAN compiler.
\psrsoft\ is tested using the GNU compiler collection (GCC\footnote{See: http://gcc.gnu.org}), however other compilers may work.
It is essential that your C and FORTRAN compilers are compatible and can cross-compile, this typically means that they must be the exact same version.
Some packages also require an Oracle/Sun compatible Java Development Kit.

\subsection{Installing and configuring \psrsoft}
\psrsoft\ is available from {\tt http://www.pulsarastronomy.net/pulsar/software/psrsoft} and is packaged as a gziped\footnote{See: {\tt http://www.gzip.org/}} {\sc tar}\footnote{See: {\tt http://www.gnu.org/software/tar/}} file.
Installing \psrsoft\ simply requires unpacking the {\sc tar} file, typically using a command such as:
\begin{verbatim}
 $ tar -xzf psrsoft.tar.gz
 $ cd psrsoft
 $ ls
bin/  config/
\end{verbatim}

\subsubsection{Configuring \psrsoft}
Let us assume that you have unpacked the \psrsoft\ install file as above.
\psrsoft\ is configured through a settings file, {\tt config/profile}, which defines variables using the {\sc bash} syntax.
An example file is included which contains default settings that will likely work for your system.
Begin by copying the example file:
\begin{verbatim}
 $ cp ./config/profile.example ./config/profile
\end{verbatim}

Most users will not need to change any settings in the profile file, however you can open it with any text editor to modify settings such as the
preferred C or Fortran compilers.
It is important to note that the C and Fortran compilers must be of the same version, typically {\tt gcc} and {\tt gfortran}.
You can check that the software versions are the same by confirming that the output of the following two commands is identical.
\begin{verbatim}
 $ gcc -v
Using built-in specs.
Target: i486-linux-gnu
Configured with: ../src/configure  ...
Thread model: posix
gcc version 4.3.2 (Debian 4.3.2-1.1)

 $ gfortran -v
Using built-in specs.
Target: i486-linux-gnu
Configured with: ../src/configure ...
Thread model: posix
gcc version 4.3.2 (Debian 4.3.2-1.1)
\end{verbatim}

\subsection{Building Pulsar Software with \psrsoft}
\psrsoft\ is invoked with the command {\tt ./bin/psrsoft}, taking a list of program names to install as arguments.
A complete listing of software that are available can be found using the {\tt --search} option.
\begin{verbatim}
 $ ./bin/psrsoft --search
==== PSRSOFT version 1.5 ====
 Pkg Index: 'stable'
=============================
Updating package index
Getting latest package descriptions... done
Searching for package .* in stable tree
1) psrchive 13.4+
2) cfitsio 3090
3) fftw 3.1.2
4) presto-core 1.01
...
\end{verbatim}

If one piece of software relies on another, \psrsoft\ will automatically download and install the dependencies before installing the requested software.
To install all the software used in the example in Section 3, we can run the following.
\begin{verbatim}
./bin/psrsoft sixproc dspsr psrchive pulsarhunter
==== PSRSOFT version 1.5 ====
 Pkg Index: 'stable'
=============================
Analysing dependancies

Packages to be installed...
===========================

  1         cfitsio |NO| (3090 2009-10-06 14:47)
  2            fftw |N | (3.1.2 2011-01-18 14:39)
  3          pgplot |N | (5.2 2012-02-10 09:37)
  4         sixproc |N | (5.0.5 2011-04-07 09:49)
  5          tempo2 |NO| (2012.1 2012-02-10 09:45)
  6           tempo |N | (11.010_sf 2011-02-12 12:30)
  7          psrxml |NO| (1.05 2010-02-04 12:30)
  8        psrchive |N | (13.4+ 2012-04-11 11:01)
  9           dspsr |N | (2.0 2012-04-12 14:52)
 10    pulsarhunter |N | (1.3r79 2011-02-23 15:21)
Install 10 packages into psrsoft/usr? (y/n)
\end{verbatim}

The 10 packages are made up of the four packages we requested, plus the dependencies.
The character {\tt N} indicates that the package is new ({\tt U} would indicate that the package needs to be updated) and the symbol {\tt O} implies that the package is optional.
For instance, if you do not require {\sc psrfits} support, you can use the {\tt --no-cfitsio} option to skip building {\tt cfitsio}.
Otherwise, answering `{\tt y}' to the prompt will begin the download and install of the requested packages.

\section{Example Pulsar Search}
For this example we will download the file {\tt S13405\_1.sf}, which has source name G123070 from the Parkes Data Archive \cite{khm+12}.
Although this is a normal survey observation, it was pointed close to the very bright pulsar PSR J0835--4510 which should be easily detected by our search.

\subsection{Extract Data}
The downloaded data normally need to be converted to a format that is understood by the search software that is to be used.
We will use the {\tt filterbank} command from {\sc sigproc}\footnote{Here we use a modified version of {\sc sigproc} available from {\tt https://github.com/SixByNine/sigproc}, or through the \psrsoft\ package `{\tt sixproc}'} to covert the downloaded data into a {\sc sigproc} compatible format.
\begin{verbatim}
 $ tar -xvf PulsarObs.tar
atnf-attribution.txt
S13405_1.sf

 $ filterbank S13405_1.sf > S13405_1.fil
\end{verbatim}

We can check the details of our data by using the {\tt header} command.
\begin{verbatim}
 $ header S13405_1.fil
Data file                        : S13405_1.fil
Header size (bytes)              : 352
Data size (bytes)                : 17154048
Data type                        : filterbank (topocentric)
Telescope                        : Parkes
Datataking Machine               : ?????
Source Name                      : G123070
Source RA (J2000)                : 08:32:37.0
Source DEC (J2000)               : -44:51:00.0
Frequency of channel 1 (MHz)     : 452.062500
Channel bandwidth      (MHz)     : -0.125000
Number of channels               : 256
Number of beams                  : 1
Beam number                      : 1
Time stamp of first sample (MJD) : 48664.615312499998
Gregorian date (YYYY/MM/DD)      : 1992/02/12
Sample time (us)                 : 300.00000
Number of samples                : 536064
Observation length (minutes)     : 2.7
Number of bits per sample        : 1
Number of IFs                    : 1
\end{verbatim}

\subsection{Dedispersion}
Dedispersion is performed using the {\tt dedisperse\_all} command.
This dedisperses to a large number of DM trials at once, with the DM range specified with the {\tt -d} option.
Since the input data file may be large, the software reads a block of data at a time, the size of which is specified with the {\tt -g} option.
Here we dedisperse from 0 to 500 cm$^{-3}$pc using a block size of 100000 samples.
\begin{verbatim}
 $ dedisperse_all S13405_1.fil -d 0 80 -g 100000
...
\end{verbatim}
This will produce a large number ($\sim 326$) files with the `{\tt .tim}' extension, each of which is a single time-series de-dispersed at a particular value of DM.

\subsection{Periodicity Search}

We can continue to use {\sc sigproc} to do the periodicity analysis with the {\tt seek} command.
This program reads in one of our time-series and performs the Fourier transform and harmonic summing, outputting a list of periodicities and S/Ns.
We can run {\tt seek} on a single tim file,
\begin{verbatim}
 $ seek S13405_1.fil.0000.00.tim -fftw -head
...
\end{verbatim}
which has output a {\tt .prd} file that contains a short header followed by columns of S/N and period in milliseconds.
Each pair of columns is a different harmonic fold, which by default is 1,2,4,8 and 16.
The period of PSR J0835--4510 is 89~ms, and the pulsar is so bright that even in the 0-DM timeseries it should be the strongest candidate with a S/N of greater than 30.

In practice we can run seek on all our {\tt .tim} files using a simple shell script.
In {\sc bash} we can run
\begin{verbatim}
 $ for timfile in S13405_1.fil.*.tim ; do seek $timfile -fftw -head ; done
 $ cat S13405_1.fil.*.prd > S13405_1.prd
\end{verbatim}
which will process all of the files and concatenate the output into a single file that we can use to identify candidates to begin the optimisation process.

\subsection{Candidate Matching}
Our combined {\tt .prd} file contains more than 10000 signals, which is clearly far too many to analyse further, however many have the same period at different DM trials, or are harmonics of other stronger signals.
We can reduce the number of candidates to something more manageable by combining signals with the same period or harmonics of the same period.
For this we can use the {\tt best} command from {\sc sigproc}, however {\sc pulsarhunter} provides more advanced matching and output options through {\tt ph-best}.

\begin{verbatim}
 $ ph-best S13405_1.prd S13405_1 --maxresults 50
...
 $ head -n 3 S13405_1.lis
S13405_1_001.phcx.gz    445.600000       89.29830545    69.324158       ...
S13405_1_002.phcx.gz    61.700000       267.79276611    57.694286       ...
S13405_1_003.phcx.gz    44.500000       290.20468989    69.324158       ...
\end{verbatim}

The main output of {\tt ph-best} is a text file with the {\tt .lis} extension.
This has columns of: candidate name; S/N; period (ms); DM (cm$^{-3}$pc); acceleration (m\,s$^{-2}$); jerk (m\,s$^{-3}$); number of harmonics matched; and best harmonic fold.
The top candidate should have a S/N of around 450, period of $\sim 89.29$~ms and DM of $\sim 69$~cm$^{-3}$pc.

In addition, {\tt ph-best} produces candidate files with the {\tt .phcx.gz} extension, which contain all the relevant information about the candidate in an XML format.

\subsection{Folding and Optimisation}
Once we have reduced the number of candidates to a reasonable number, we can fold the signals to produce diagnostic plots useful for identifying the candidates that are suitable for follow-up at the telescope.
We can fold the data using {\sc dspsr} on the original {\tt  .fil} file.
{\sc dspsr} can fold many signals at once, however for this example we will fold each candidate in turn.
We can specify the folding period, in seconds, using the {\tt -c} option, and the DM using the {\tt -D} option; we can obtain these values for each candidate from the .lis file.
Other useful options are {\tt -L} to specify a sub-integration time (in seconds), {\tt -t} to fold using multiple threads, {\tt -e} to specify the output extension and {\tt -U} to limit the memory usage to a value small enough to fit within the CPU cache.

\begin{verbatim}
 $  dspsr S13405_1.fil -c 0.08929830545 -D 69.324158 -L 10 -t 4 -U 1 -e subint
...
\end{verbatim}

This produces a number of {\tt .subint} files, which we can add together using {\tt psradd} from {\sc psrchive}.
It is also useful to reduce the number of frequency channels to, for instance, 8.

\begin{verbatim}
 $ psradd *.subint -o S13405_1_001.ar
 $ rm *.subint
 $ pam --setnchn 8 -m S13405_1_001.ar
...
\end{verbatim}

Now that we have a {\tt psrchive} archive of our pulsar file, we can use any of the {\tt psrchive} tools discussed in van Straten \& Demorest (in prep).
In particular we want to use {\tt pdmp} to optimise the S/N of the candidate by fine-tuning the period and DM.
This also produced the diagnostic plots that are useful for distinguishing pulsars from RFI and noise.
The {\tt phcx.gz} files we created can also be updated with the additional information from {\tt pdmp}, however pdmp also produces a postscript output that can be quickly viewed.

\begin{verbatim}
 $ pdmp -input-phcx S13405_1_001.phcx.gz -output-phcx S13405_1_001.phcx S13405_1_001.ar
...

 $ gzip -f S13405_1_001.phcx
\end{verbatim}

Since we need to do this for each of our candidates, it is again useful to create a script to automate the process.
Below is an example {\sc bash} script to fold and optimise each of our candidates.
\begin{verbatim}
#!/bin/bash
file=S13405_1

while read line ; do
   set -- ${file}
   cand=`basename $1 .phcx.gz` # strip file extension
   psriod=`echo $2 | awk "{print $1/1000.0}" # convert to seconds
   dm=$3
   dspsr ${file}.fil -c ${period} -D ${dm} -L 10 -t 4 -U 1 -e subint
   psradd *.subint -o ${cand}.ar
   rm *.subint
   pam --setnchn 8 -m ${cand}.ar

   pdmp -input-phcx ${cand}.phcx.gz -output-phcx ${cand}.phcx ${cand}.ar
   gzip -f ${cand}.phcx
done < ${file}.lis

\end{verbatim}

\begin{figure}[bt]
\begin{center}
\includegraphics[width=14cm]{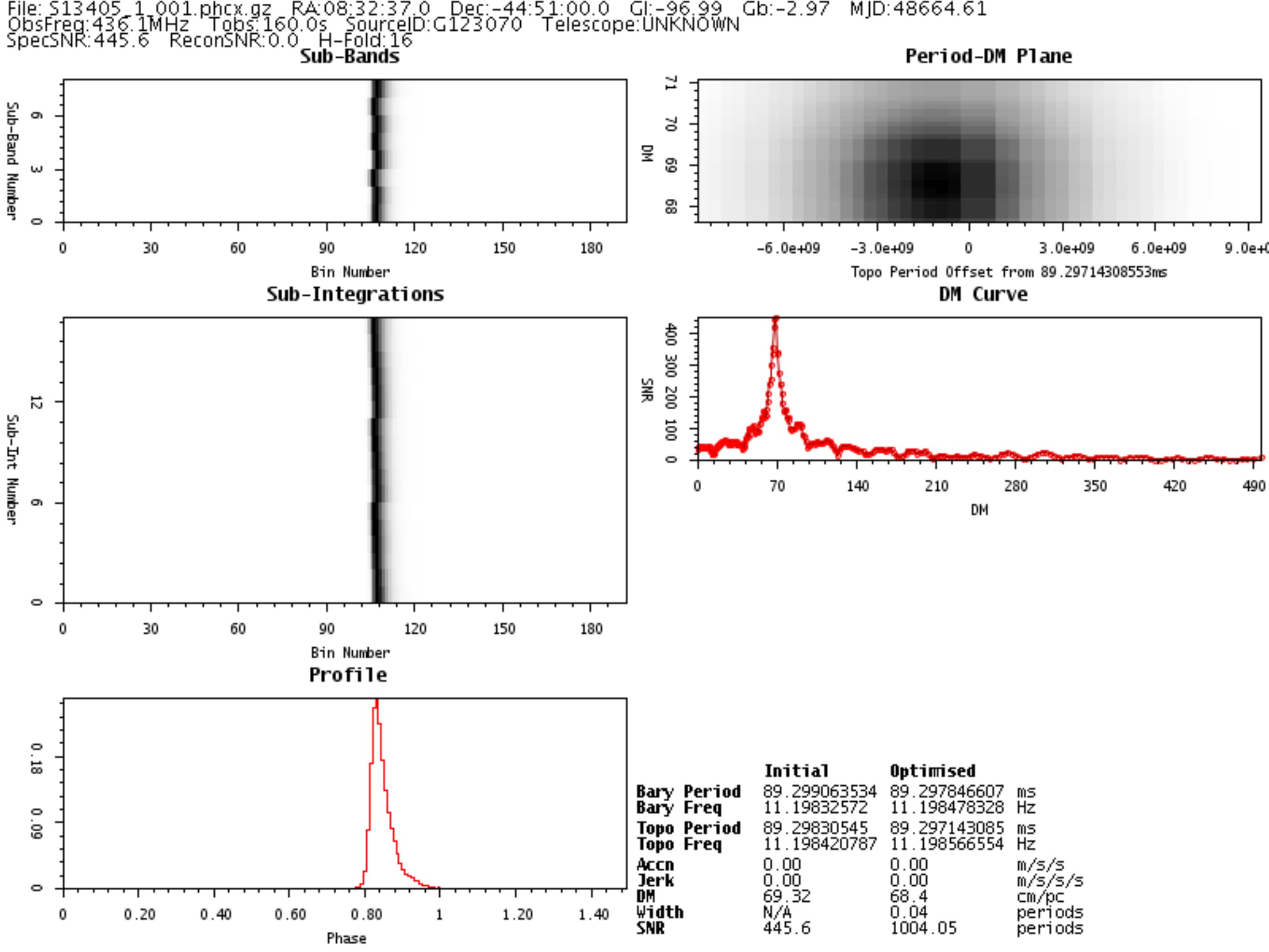}
\caption{
\label{candidateplot}
Example candidate plot showing the bright pulsar PSR J0835--4510.
Panels are, from top to bottom, on the left side: signal strength as a function of frequency channel and pulse phase; signal strength as a function of sub-integration and pulse phase; and signal strength as a function of pulse phase.
On the right side: S/N as a function of trial period and DM; and S/N as a function of DM.
}
\end{center}
\end{figure}

\subsection{Viewing Output}
The postscript output from pdmp can be viewed with any standard postscript viewer, however more information can be obtained from the {\tt .phcx.gz} files.
These files can be viewed using either the {\tt JReaper} graphical candidate selection tool \cite{kel+09}, or individually using the {\tt ph-view-phcx} command.

\begin{verbatim}
 $ ph-view-phcx S13405_1_001.phcx.gz
\end{verbatim}

or to make a png image,
\begin{verbatim}
 $ ph-view-phcx S13405_1_001.phcx.gz --imageoutput
 $ open S13405_1_001.phcx.gz.png
\end{verbatim}

The candidate plot for our example candidate {\tt S13405\_1\_001} is shown in Figure \ref{candidateplot}.
The key features are that there is a clear peak in the `DM curve', i.e. S/N as a function of DM and the pulsar is visible in all (or most) sub-integrations and sub-bands.
Identifying pulsars and distinguishing them from RFI can be a difficult task, however this becomes easier with experience.

If a good candidate pulsar is identified, it is recommended to ensure that it not already known by consulting the ATNF pulsar catalogue\footnote{http://www.atnf.csiro.au/research/pulsar/psrcat/}.
If the candidate is not a harmonic of a known pulsar, then confirmation of the pulsar can be done with follow-up observations, or by searching previous observations of the same location in archival data.

\section{Conclusion}
Although there are many different ways to search for radio pulsars, most common techniques follow the same steps as described here.
Utilising data from different telescopes can bring new challenges, however most data formats are relatively simple and can easily be converted.
In future, it is likely that all telescopes will generate data in the same {\sc psrfits} format as used by the Parkes Data Archive, which is readable by most pulsar search software.
Therefore, by following the principals in this paper, you should be able to process any pulsar search data available.

\begin{acknowledgements}
MJK would like to acknowledge the many people responsible for creating and maintaining the software described in this tutorial.
In particular, to Duncan Lorimer for originally creating {\sc sigproc}; Matthew Bailes for creating {\sc dedisperse\_all}; to Willem van Straten, Aiden Hotan, Andrew Jameson and Paul Demorest for {\sc psrchive} and {\sc dspsr}; and to Dick Manchester for designing and maintaining the {\sc psrfits} standard.
\end{acknowledgements}

\appendix                  

\section{Example Processing Script}
This appendix contains a {\sc bash} script that will process a provided {\sc psrfits} format file.
Readers can download from http://zmtt.bao.ac.cn/psr/soft/processPSRFITS.csh
\label{lastpage}

\end{document}